\pdfoutput=1

\documentclass[conference]{IEEEtran}
\IEEEoverridecommandlockouts
\usepackage{cite}
\usepackage{amsmath,amssymb,amsfonts}
\usepackage{algorithmic}
\usepackage{graphicx}
\usepackage{textcomp}
\usepackage{xcolor}
\def\BibTeX{{\rm B\kern-.05em{\sc i\kern-.025em b}\kern-.08em
    T\kern-.1667em\lower.7ex\hbox{E}\kern-.125emX}}
    
    \allowdisplaybreaks
\begin{document}

\title{Short-term Operational Planning Problem of the Multiple-Energy Carrier Hybrid AC/DC Microgrids}

\author{
\IEEEauthorblockN{1\textsuperscript{st} Reza Bayani}
\IEEEauthorblockA{\textit{\small Department of Electrical and Com. Eng.} \\
\textit{San Diego State University}\\
San Diego, USA 92182 \\
rbayani@sdsu.edu}
\and
\IEEEauthorblockN{2\textsuperscript{nd} Mohammed Bushlaibi}
\IEEEauthorblockA{\textit{\small Department of Electrical and Com. Eng.} \\
\textit{San Diego State University}\\
San Diego, USA 92182  \\
mbushlaibi3272@sdsu.edu}
\and
\IEEEauthorblockN{3\textsuperscript{nd} Saeed D. Manshadi}
\IEEEauthorblockA{\textit{\small Department of Electrical and Com. Eng.} \\
\textit{San Diego State University}\\
San Diego, USA 92182 \\
smanshadi@sdsu.edu}}

\maketitle

\begin{abstract}
In this paper, the short-term operation problem for a multiple energy carrier hybrid AC/DC microgrid is discussed. The hybrid microgrid consists of AC and DC parts, which are connected by means of inverters as well as natural gas network. The microgrid includes photovoltaic (PV) unit, wind turbine (WT), battery storage unit and gas-fired microturbines. A mixed integer linear programming is formed to minimize the overall cost of the microgrid including cost of natural gas supply, the value of lost load and battery degradation cost. The presented case study explored the importance of inverter characteristics and pipeline capacity.
\end{abstract}


\section*{Nomenclature}
\subsection*{Variables}
\noindent \begin{tabular}{ l p{6.55cm} }
$E$ & Available energy of battery storage unit\\
$f,\pi,v$ & Natural gas flow/pressure/supplier output\\
$P,Q$ & Active/Reactive power dispatch\\
$PL,QL,SL$ & Active/Reactive/Apparent power of line\\
$V,\theta$ & Voltage magnitude and angle\\
\end{tabular}
\subsection*{Indices}
\noindent \begin{tabular}{ l p{6.55cm} }
$c$ & Inverter\\
$ch,dc$ & Charge/Discharge of the battery\\
$d$ & Demand served\\
$F$ & Forecast value of renewable unit output\\
$g,gs$ & Microturbine/Supplier\\
$j,o,m,n$ & Energy hub\\
$k$ & Battery unit\\
$p$ & Natural gas pipe\\
$s,w$ & Solar/Wind turbine unit\\
$t$ & Time\\
\end{tabular}
\subsection*{Parameters}
\noindent \begin{tabular}{ l p{6.55cm} }
$c_{p}$ & Pipeline constant\\ 
$C_{gs}()$ & Cost function of natural gas supplier\\
$F_g()$ & Fuel consumption function of microturbine $g$\\
$G_{j,o},B_{j,o}$& Real/Imaginary part of admittance matrix\\
$\beta$ & Degradation cost of battery\\   
$\eta$ & Charging/discharge efficiency of the battery\\
$\kappa_e,\kappa_g$ & Value of lost load for electricity/gas demand\\
$\pi {}_{n}'$& Initial natural gas pressure at energy hub n\\
\end{tabular}

\section{Introduction}
At present, the majority of the electricity generation in the United States is provided by natural gas fueled units \cite{eia2020}. It is also estimated that in the next 30 years, natural gas will remain the top contributing fuel for electricity generation. A detailed examination of the simultaneous operation of the natural gas and electricity grid is presented in \cite{correa2014integrated}. These facts, in combination with the concept of microgrids as a means of ensuring transition towards a sustainable power network in the future, justify the promotion of the multiple energy carrier microgrid concept.  A multiple energy carrier microgrid is an entity which consists of distributed energy resources (DERs) and demand, and is able to operate independent of the main grid \cite{manshadi2015resilient}. In this setup, natural gas and electricity demands are served within each energy hubs, which are coupled with electricity and natural gas networks. \\
There is push toward usage of sources as well as demand with DC power which promoted the idea of utilizing a DC grid. In comparison with AC networks, DC networks are more efficient, don't have grid synchronization concerns, and are less affected by utility side disturbances. There are a number of works that address the hyrbid AC/DC microgrids. By taking the advantages of both AC and DC frameworks, a short-term operation framework for hybrid AC/DC microgrids is presented in \cite{manshadi2016decentralized}.  A power control strategy using multiple inverters for an hybrid AC/DC microgrid system is proposed in \cite{ambia2011centralized}. An optimal planning model is proposed in \cite{lotfi2016static} to identify the the minimum planning cost of DERs, inverters, energy exchange with the utility grid, and the cost of the unused energy. By analyzing energy efficiency and cost effectiveness of a renewable integrated network, a distribution planning strategy for a hybrid AC/DC system is performed in \cite{tian2020cost}. An optimal scheduling for a hybrid AC/DC microgrid is proposed in \cite{alanazi2017coordinated} in order to minimize the operation cost of the network.\\
In this paper, the concept of multiple energy carrier hybrid AC/DC microgrid is presented. This system not only benefits from the quality and the reliability of the energy provision offered by a multiple-energy carrier system, but also enjoys the perks of a hybrid AC/DC system. This work formulates the short-term operation of the proposed setup. The presented case study explored the impact of some physical limitation and operational charactristics under various operating conditions.\\
\section{Problem formulation and solution methodology}
The operational requirements of the multiple-energy carrier microgrid which has electricity and natural gas networks with three constraint groups, which are individually discussed in this section. Then the objective of multiple-energy carrier hybrid AC/DC microgrid is presented. \subsection{The AC Electricity Network Operating Constraints}
The electrical network constraints for the AC side are presented in \eqref{eq:ac_constraints}. The lower and upper band limits for the voltage magnitude of each energy hub, and active and reactive power output of the micro-turbines are given in \eqref{eq:voltage_limit}, \eqref{eq:p_gen_limit}, and \eqref{eq:q_gen_limit}, respectively. The dispatched power of the the wind turbine and the solar units at each hour are presented in \eqref{eq:wind_forecast} and \eqref{eq:solar_forecast}, respectively. By \eqref{eq:q_wind} and \eqref{eq:q_solar}, it is considered that the inverters placed on each renewable unit are able to produce positive and negative reactive powers, with an absolute value of less than the dispatched active power of each unit. The nodal active and reactive power balance at each hour of the day for each energy hub are presented in \eqref{eq:ac_balance_active} and \eqref{eq:ac_balance_reactive}, respectively.
\begin{subequations} \label{eq:ac_constraints}
\begin{align}
& V^{min} \leq V_j^t \leq V^{max} \label{eq:voltage_limit}\\
& P^{min}_g \leq P_{g,t} \leq P^{max}_g \label{eq:p_gen_limit}\\
& Q^{min}_g \leq Q_{g,t} \leq Q^{max}_g \label{eq:q_gen_limit}\\
& 0 \leq P_{w,t} \leq P^F_{w,t} \label{eq:wind_forecast}\\
& 0 \leq P_{s,t} \leq P^F_{s,t} \label{eq:solar_forecast}\\
& - P_{w,t} \leq Q_{w,t} \leq P_{w,t}
\label{eq:q_wind}\\
& -P_{s,t} \leq Q_{s,t} \leq P_{s,t} \label{eq:q_solar}\\
&\begin{aligned}
\sum _{g\in D_{j_{ac}}^{g}} P_{g,t}+\sum _{w\in D_{j_{ac}}^{w}} P_{w,t} + \sum_{s \in D_{j_{ac}}^{s}} P_{s,t}\\-\sum _{c\in D_{j_{ac}}^{c}} P_{c,t} -P_{j_{ac},t}^{d}={P_{j_{ac},t}^{inj}}
\label{eq:ac_balance_active}\end{aligned}\\
&\begin{aligned}
\sum _{g\in D_{j_{ac}}^{g}} Q_{g,t}+\sum _{w\in D_{j_{ac}}^{w}} Q_{w,t} + \sum_{s \in D_{j_{ac}}^{s}} Q_{s,t}\\-\sum _{c\in D_{j_{ac}}^{c}} Q_{c,t} -Q_{j_{ac},t}^{d}={Q_{j_{ac},t}^{inj}}
\label{eq:ac_balance_reactive}\end{aligned}\\
&\begin{aligned}
P_{j_{ac},t}^{inj}=(2V_{j}^{t}-1)G_{j,j} \\
+\sum_{o(o\neq j)}{(G_{j,o}(V_{j_{ac}}^{t}}+V_{o_{ac}}^{t}-1)
+B_{j,o}(\theta_{j_{ac}}^{t}-\theta_{o_{ac}}^{t}))
\label{eq:ac_inject_active}\end{aligned}\\
&\begin{aligned} 
Q_{j_{ac},t}^{inj}=(-2V_{j}^{t}-1)B_{j,j}\\
+\sum_{o(o\neq j)}{(G_{j,o}(\theta _{j_{ac}}^{t}-\theta  _{o_{ac}}^{t})-B_{j,o}(V_{j_{ac}}^{t}}{+V_{o_{ac}}^{t}-1))}
\label{eq:ac_inject_reactive}\end{aligned}\\
& PL_{j_{ac,}o_{ac}}^{ac,t}=-G_{j,o}(V_{j_{ac}}^{t}-V_{o_{ac}}^{t})+B_{j,o}(\theta_{j_{ac}}^{t}-\theta _{o_{ac}}^{t})
\label{eq:ac_line_active}\\
& QL_{j_{ac,}o_{ac}}^{ac,t}=B_{j,o}(V_{j_{ac}}^{t}-V_{o_{ac}}^{t})+G_{j,o}(\theta_{j_{ac}}^{t}-\theta _{o_{ac}}^{t})\label{eq:ac_line_reactive}\\
& SL_{j,o_{ac}}^{ac,t}=PL_{j,o_{ac}}^{ac,t}+\xi \cdot QL_{j,o_{ac}}^{ac,t}\label{eq:ac_line_apparent}\\
& |SL_{j,o_{ac}}^{ac,t}| \leq SL_{j,o_{ac}}^{max}\label{eq:ac_limit_apparent}\\
& P_{c}^{min}\leq P_{c,t}\leq P _{c}^{max}\label{eq:inverter_limit_active}\\
& Q_{c}^{min}\leq Q_{c,t}\leq Q_{c}^{max}\label{eq:inverter_limit_reactive}
\end{align}
\end{subequations}
The net injected active and reactive power at each energy hub of the AC network are calculated according to \eqref{eq:ac_inject_active} and \eqref{eq:ac_inject_reactive}, respectively. The active and reactive power flow within the power line in the AC network are calculated based on \eqref{eq:ac_line_active} and \eqref{eq:ac_line_reactive}. The apparent power of each line is then obtained based on \eqref{eq:ac_line_apparent}. Here, $\xi$ is a parameter which is related to the power factor of the demand. This kind of approximation for apparent power calculation of the line can be found in other references \cite{manshadi2016decentralized}. Finally, the limit for absolute value of each line's apparent power is enforced by \eqref{eq:ac_limit_apparent} and the limits for the active and reactive power injected from the AC/DC inverter are given by \eqref{eq:inverter_limit_reactive} and \eqref{eq:inverter_limit_reactive}.
\subsection{The DC Electricity Network Operating Constraints}
The operating constraints for the DC side of the electrical network are according to \eqref{eq:dc_constraints}. Equations \eqref{eq:bat_charge} and \eqref{eq:bat_disch} enforce the upper and lower bound limits of the battery storage system charging and discharging at all times. Based on \eqref{eq:bat_binary}, no simultaneous charging and discharging can happen. The available energy in the storage unit at each hour is obtained based on \eqref{eq:bat_ch_dc}. The minimum and maximum limits for the stored energy in battery storage system at each hour are shown in \eqref{eq:bat_e_limit}. The active power transmitted through the DC line is given by \eqref{eq:dc_line_active}. The active power nodal balance for each energy hub of the DC network is shown in \eqref{eq:dc_balance}. It is noteworthy that the inverter active power is assigned opposite signs in the active power balance equations for AC and DC sides.
\begin{subequations} \label{eq:dc_constraints}
\begin{alignat}{3}
& I_{ch,k,t}\cdot P^{min}_{ch,k,t}\leq P_{ch,k,t}\leq I_{ch,k,t}\cdot P^{max}_{ch,k,t} \label{eq:bat_charge}\\
& I_{dc,k,t}\cdot P^{min}_{dc,k,t}\leq P_{dc,k,t}\leq I_{dc,k,t}\cdot P^{max}_{dc,k,t} \label{eq:bat_disch}\\
& I_{ch,k,t} + I_{dc,k,t} \leq 1 \label{eq:bat_binary}\\
& E_{k,t}= E_{k,t-1}-(\frac{P_{dc,k,t}}{\eta _{dc}^{k}}-\eta _{ch}^{k} \cdot P_{ch,k,t}) \label{eq:bat_ch_dc}\\
& E_{k}^{min}\leq E_{k,t}\leq E_{k}^{max}\label{eq:bat_e_limit}\\
& PL_{j_{dc,}o_{dc}}^{dc,t}=(\frac{V_{j_{dc,}t}-V_{o_{dc,}t}}{R_{dc,j_{ac}o_{ac}}})\label{eq:dc_line_active}\\
& -SL_{j,o_{dc}}^{max}\leq PL_{j,o_{dc}}^{dc,t} \leq SL_{j,o_{dc}}^{max}\label{eq:dc_limit_apparent}\\
&\begin{aligned}\sum _{k\in D_{j_{dc}}^{k}}(P_{dc,k,t}-P_{ch,k,t})+\sum _{g\in D_{j_{dc}}^{g}} P_{g,t}+\sum _{w\in D_{j_{dc}}^{w}} P_{w,t}\\ + \sum_{s \in D_{j_{dc}}^{s}} P_{s,t}
+\sum _{c\in D_{j_{dc,t}}^{c}}P_{c,t}-P_{j_{dc,}t}^{d}=\sum _{o_{dc}}PL_{j_{dc,}o_{dc}}^{dc,t}
\label{eq:dc_balance}\end{aligned}
\end{alignat}
\end{subequations}
\subsection{Natural Gas Network constraints}
The natural gas network constraints are displayed in  \eqref{eq:natural_gas_constraints}. The minimum and the maximum amount of natural gas provided by the supplier at each time is presented in \eqref{eq:supplier_limit}. The lower and upper bound for natural gas pressure at each energy hub are given in \eqref{eq:pressure_limit}. The absolute value for natural gas flow rate at each pipeline is limited according to \eqref{eq:flow_limit}. The served natural gas demand at each energy hub is displayed in \eqref{eq:gas_dispatch}. The natural gas flow equation for each pipeline is obtained based on \eqref{eq:flow_equation}. This linearization technique has been successfully implemented in other researches \cite{manshadi2015resilient}. Finally, the natural gas balance equation for each energy hub is given by \eqref{eq:gas_balance}.
\begin{subequations} \label{eq:natural_gas_constraints}
\begin{alignat}{3}
& v_{gs}^{min}\leq v_{gs,t} \leq v_{gs}^{max}\label{eq:supplier_limit}\\
& \pi_n^{min}\leq \pi_{n,t}\leq\pi_n^{max}\label{eq:pressure_limit}\\
& -f_{p}^{max} \leq f_{p}^{t}\leq f_{p}^{max} \label{eq:flow_limit}\\
& 0\leq g_{n,t}^{d}\leq g_{n,t}^{D}\label{eq:gas_dispatch}\\
&\begin{aligned}
     \sum _{gs\in GS_n} v_{gs,t} -\sum _{p\in P_{from,n}}f_{p,t}+\sum _{p\in P_{to,n}} f_{p,t}\\=\sum_{g\in G_{n}} F_{g}(P_{g,t})+g_{n,t}^{d}
\label{eq:gas_balance}\end{aligned}\\
& f_{p,t}^{n,m}=c_{p}\frac{\pi_{n}'\pi_{n,t}-\pi_{m}'\pi_{m,t}}{\sqrt{\pi_{n}'^{2}- \pi_{m}'^{2}}} \label{eq:flow_equation}
\end{alignat}
\end{subequations}
\subsection{Objective Function}
The objective function for the short-term operation problem of the multiple-energy carrier hybrid AC/DC microgrid is given by \eqref{eq:gas_objective}. Here, the first term represents the natural gas provision cost. The second and the third terms penalize the lost electricity and heat loads, respectively. Lastly, the battery system's degradation cost is included.
\begin{equation}
    \begin{aligned}
         Min\sum_{t} \sum_{gs} C_{gs}(v_{gs,t})+ \kappa_{e}\sum_{t}\sum_{j}( P_{j,t}^{D}-P_{j,t}^{d}) \\+ \kappa_{g}\sum_{t}\sum_{n}( g_{n,t}^{D}-g_{j,t}^{d}) +\beta\sum_{t}\sum_{k}(P_{dc,k,t})
    \end{aligned}
 \label{eq:gas_objective}\\
\end{equation}
To circumvent unnecessary complexities of non-linear programming, piecewise linearization technique is applied for estimating the fuel consumption function of the gas-fired microturbines and the cost function of the natural gas supplier. By doing so, the minimization of objective function \eqref{eq:gas_objective}, subject to the set of the constraints \eqref{eq:ac_constraints}, \eqref{eq:dc_constraints}, and \eqref{eq:natural_gas_constraints} becomes a mixed integer linear programming (MILP) problem, which can be solved optimally by numerous solvers.
\section{Results and Discussion}
In this section the simulation results for the operation of the proposed integrated AC/DC network and the natural gas network are presented.
\subsubsection{Network structure}
\begin{figure}[t!]
    \centering
    \vspace{-0.35cm}
    \includegraphics[width=9.2cm]{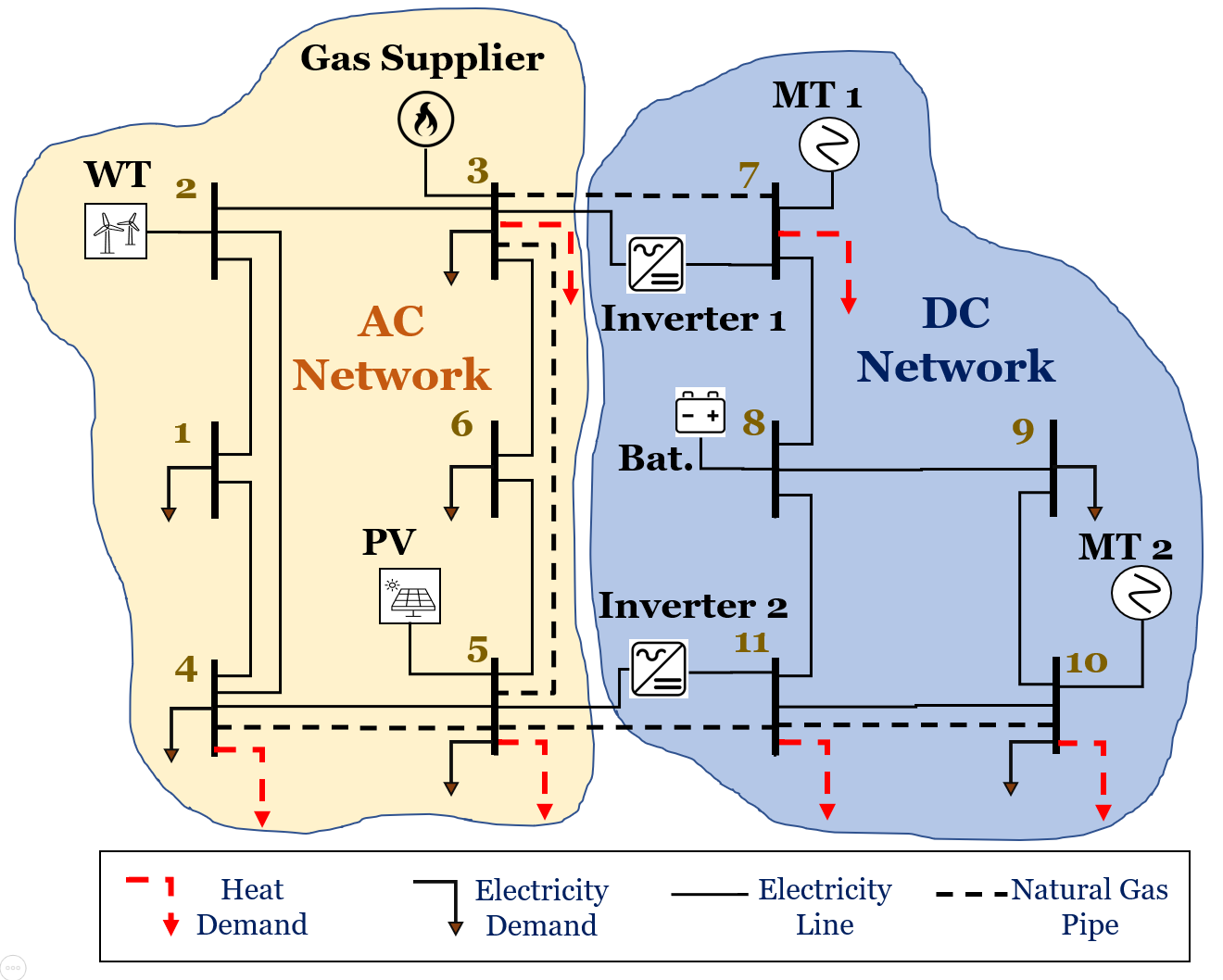}
        \vspace{-0.65cm}
    \caption{Multiple energy carrier microgrid}
    \label{fig:structure}
\end{figure}

The proposed multiple-energy carrier hybrid AC/DC microgrid is, as displayed in Fig. \ref{fig:structure}. The AC/DC electricity network consists of 12 lines, and 2 gas fired microturbines which are placed on the DC side. The AC network includes 6 energy hub, and enjoys two renewable resources, i.e. wind turbine and the solar generation unit. Two inverters are responsible for the AC/DC and DC/AC conversion of the power. The DC network includes 5 energy hubs and a storage unit.\\
The natural gas network is composed of 5 pipes which connect 6 energy hubs, and one supplier unit which is placed on hub number 3. The natural gas network has to supply two types of demands, heat and fuel. The former is the requested demand for residential usage, which is known beforehand and varies throughout the day. The latter is the fuel demand of the microturbines. It is supposed that the fuel demand must be met at all times, while heat demand can be shed. 
\subsection{Case 1 - Normal operation condition}
\begin{figure}[t!]
    \centering
    \includegraphics[width=8.0cm]{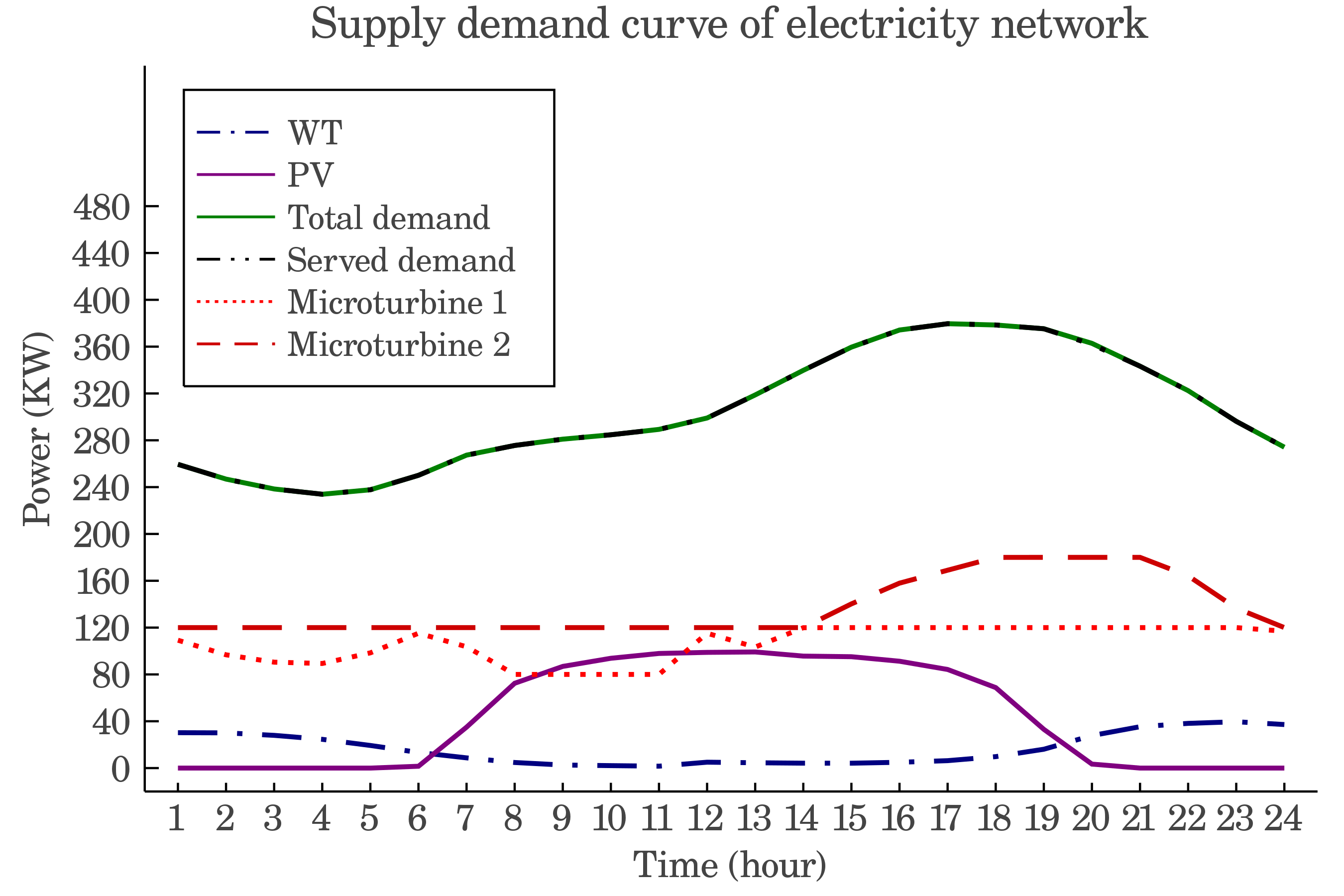}
        \vspace{-0.35cm}
    \caption{Electricity generation and consumption in Case 1}
    \label{fig:case1_elec_dispatch}
    \vspace{0.4cm}
    \includegraphics[width=8.0cm]{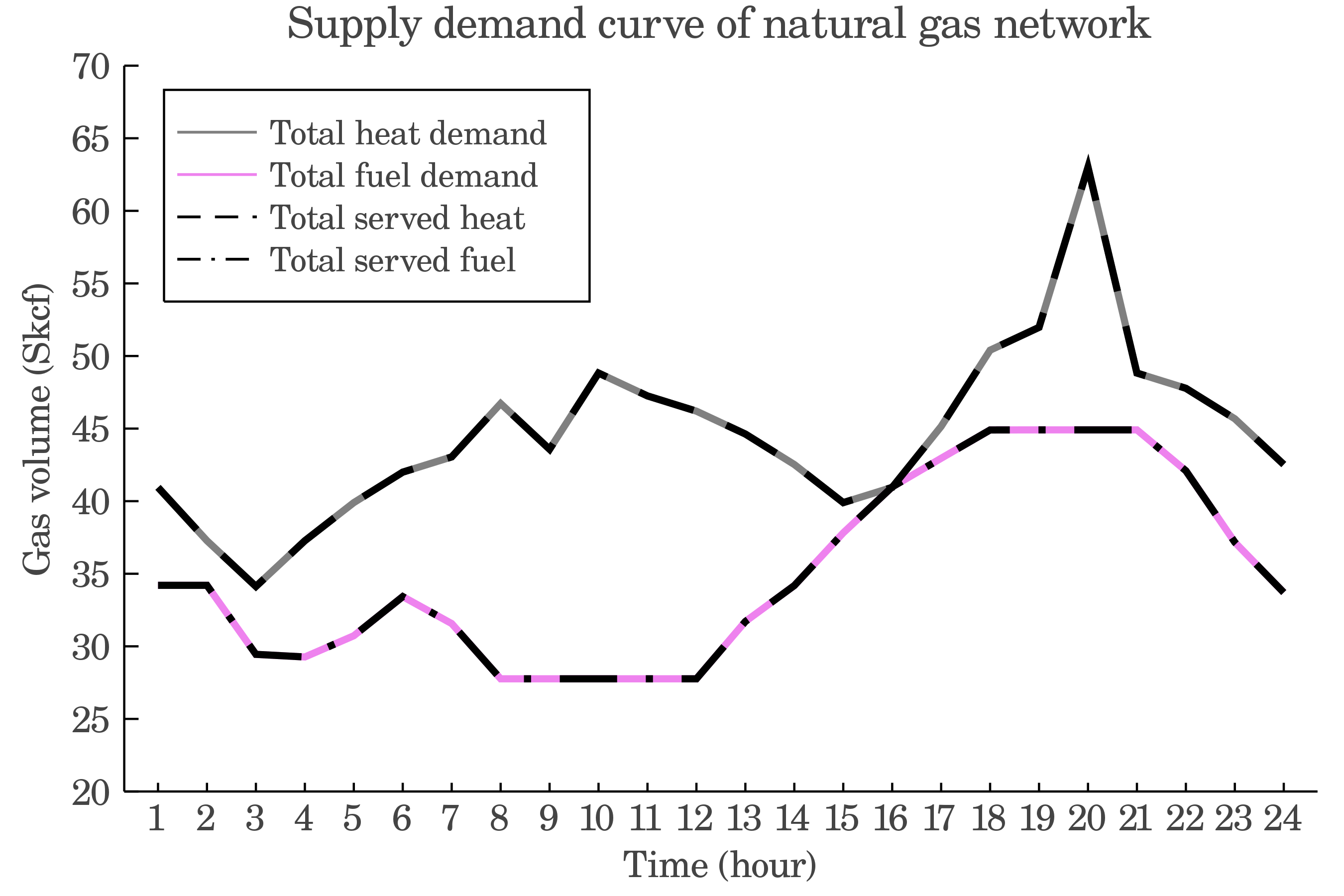}
            \vspace{-0.35cm}
    \caption{Natural gas supply demand in Case 1}
    \label{fig:case1_gas_dispatch}
    \vspace{0.4cm}
    \includegraphics[width=8.0cm]{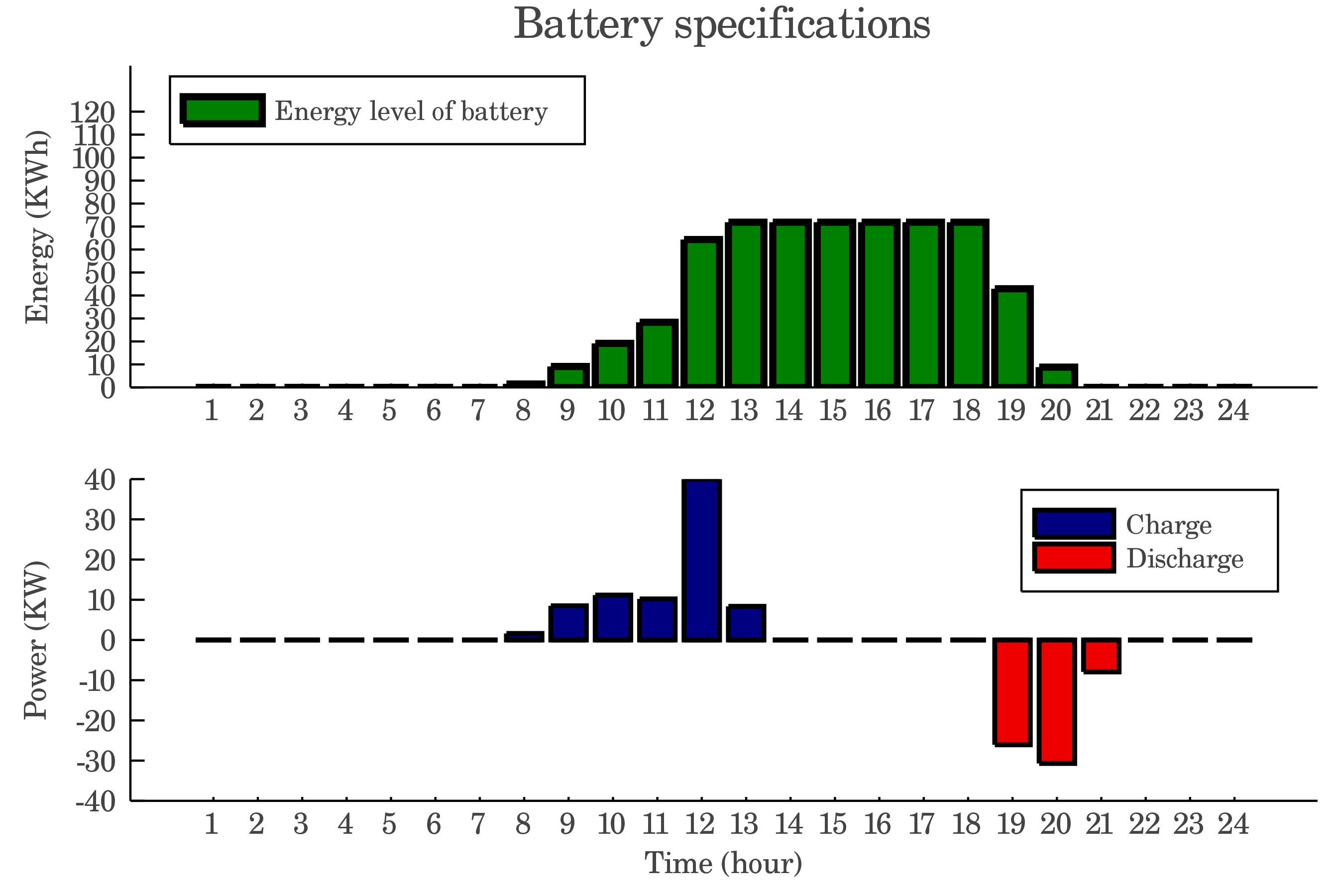}
            \vspace{-0.35cm}
    \caption{Battery energy level and power in Case 1}
    \label{fig:case1_bat}
\end{figure}

In this case, the daily operational characteristics of the proposed framework are displayed and discussed. The total generation capacity is considered equal to 300 KW, and the peak demand is equal to 380 KW. The maximum output of solar generation and wind turbine are considered equal to 100 KW and 40 KW, respectively.\\
Fig. \ref{fig:case1_elec_dispatch} shows the dispatch of different units in the electricity network, as well as the amount of served demand. The supply demand of natural gas network can also be seen in Fig. \ref{fig:case1_gas_dispatch}. It is noticed that all of the electricity demand as well as the natural gas demand is fully met. At the hour 20 of the day, both microturbines have reached their maximum power, 120 and 180 KW, respectively. The demand at this hour is equal to 362.8 KW, and the combined generation of PV and WT at this hour is equal to 31.2 KW. This deficiency is supported through the 30.7 KW discharged power of the battery system at this hour. The charge/discharge power of battery and its energy level are illustrated in Fig. \ref{fig:case1_bat}. The battery degradation cost in this case is \$144.6.\\
\subsection{Case 2 - Investigating the impact of bidirectional AC/DC inverter capacity} 
\begin{figure}[ht!]
    \centering
       \includegraphics[width=8.0cm]{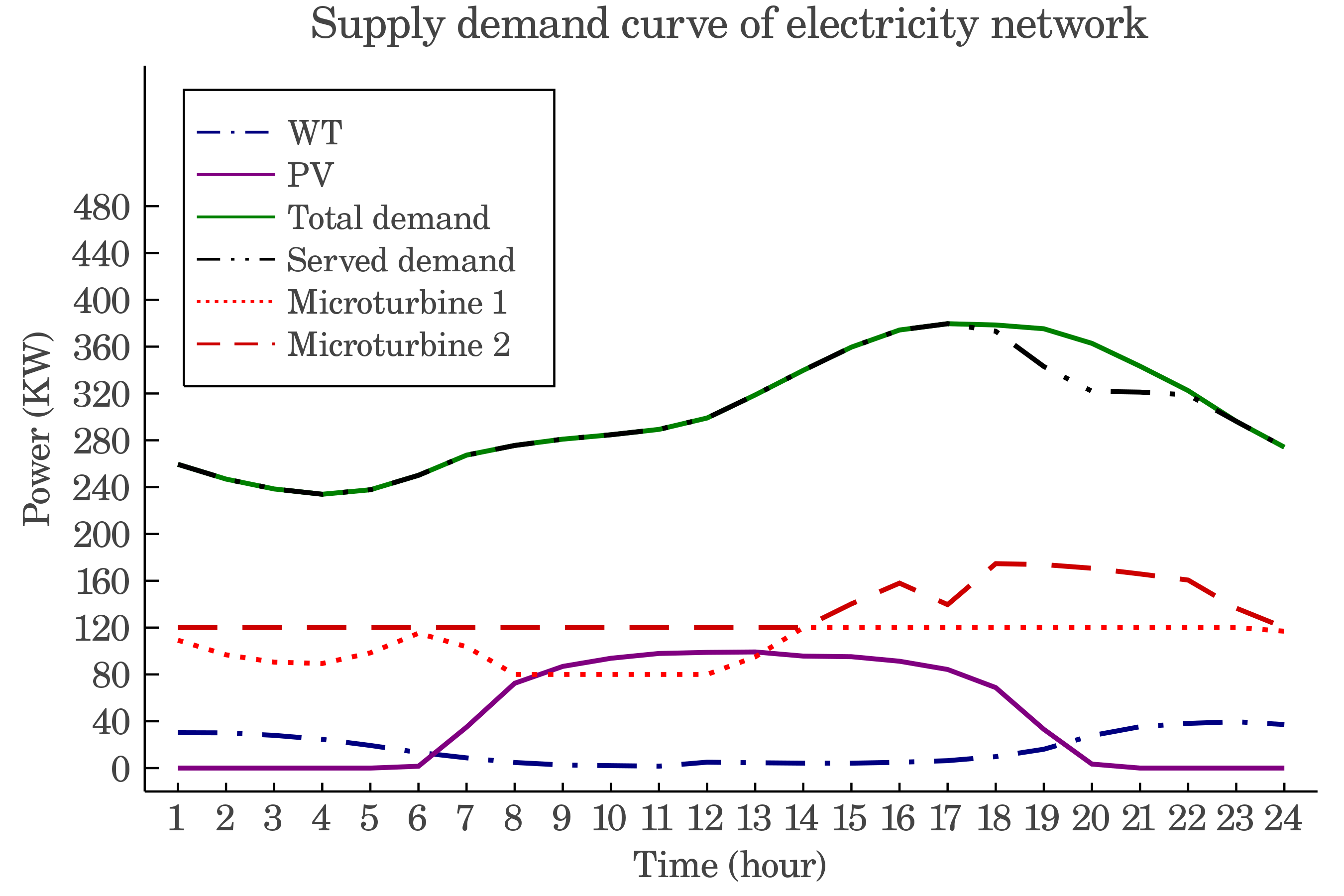}
        \vspace{-0.35cm}
    \caption{Electricity generation and consumption in Case 2}
    \label{fig:case2_elec_dispatch}
    \vspace{0.4cm}
    \includegraphics[width=8.0cm]{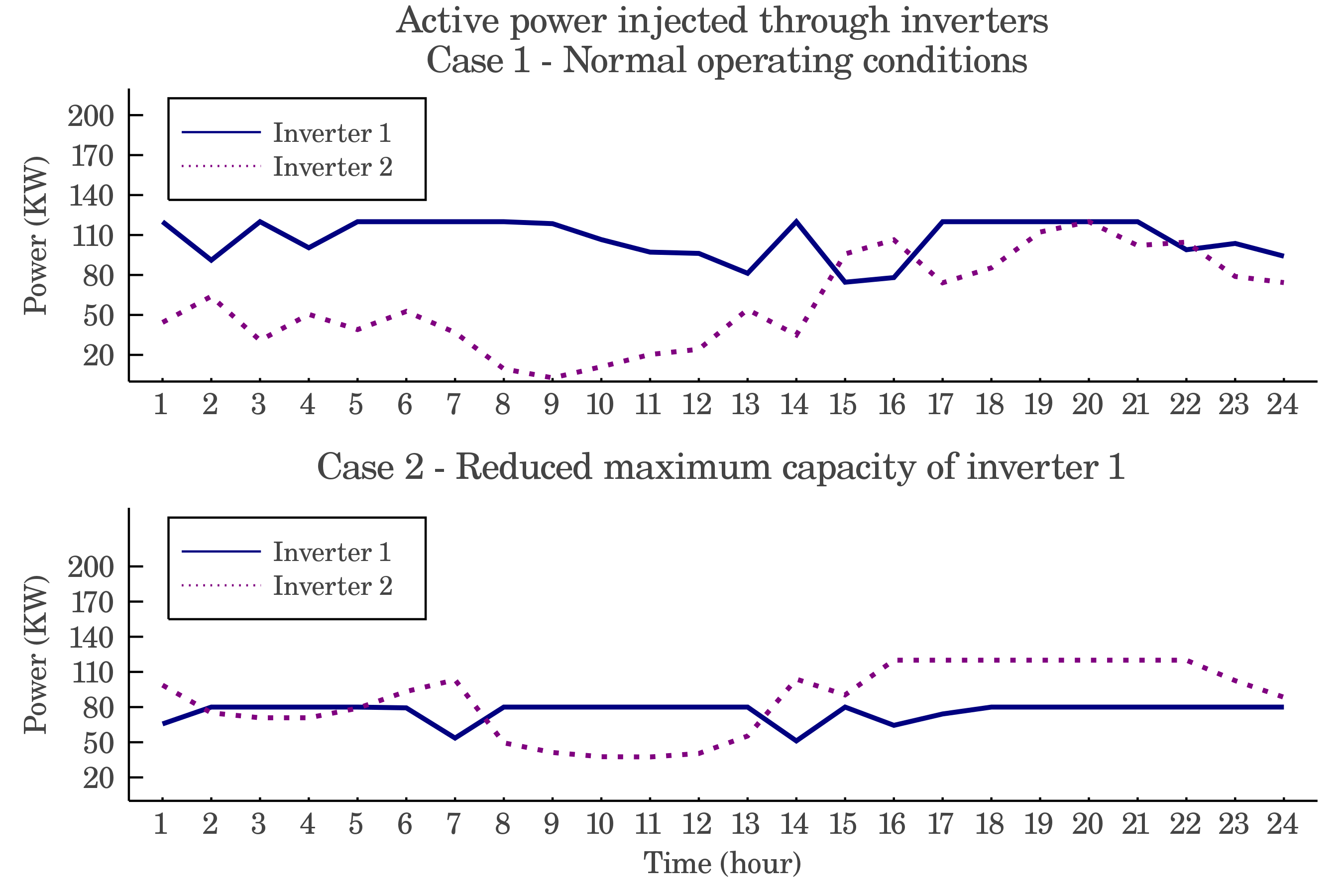}
        \vspace{-0.35cm}
    \caption{Comparison of active power injection of inverters}
    \label{fig:case2_inverter_active}
\end{figure}

To illustrate the importance of the inverters in the proposed structure, this case is designed. To this end, the maximum power capacity of inverter 1 is reduced from 120 KW to 80 KW. By doing so, the amount of power that can be injected from the DC side of the grid to the AC side is limited, and consequently, the system will fail in meeting all of the demand. As shown in Fig. \ref{fig:case2_elec_dispatch}, during hours 18-22 of the day, some demand is not served.\\
Fig \ref{fig:case2_inverter_active} shows the inverter power in cases 1 and 2, respectively. As seen in this figure, the two inverters never reached their maximum power simultaneously in case 1. However, in case 2, it is observed that during hours 18-22 of the day, both of the inverters are operating at their maximum power ratings, and thus no more power can be sent to the AC side to meet the load. As a result, 104.2 KWh of demand energy is missed during this period.
\subsection{Case 3 - Investigating the impact of natural gas pipeline capacity}
\begin{figure}[h!]
    \centering
    \includegraphics[width=6.8cm]{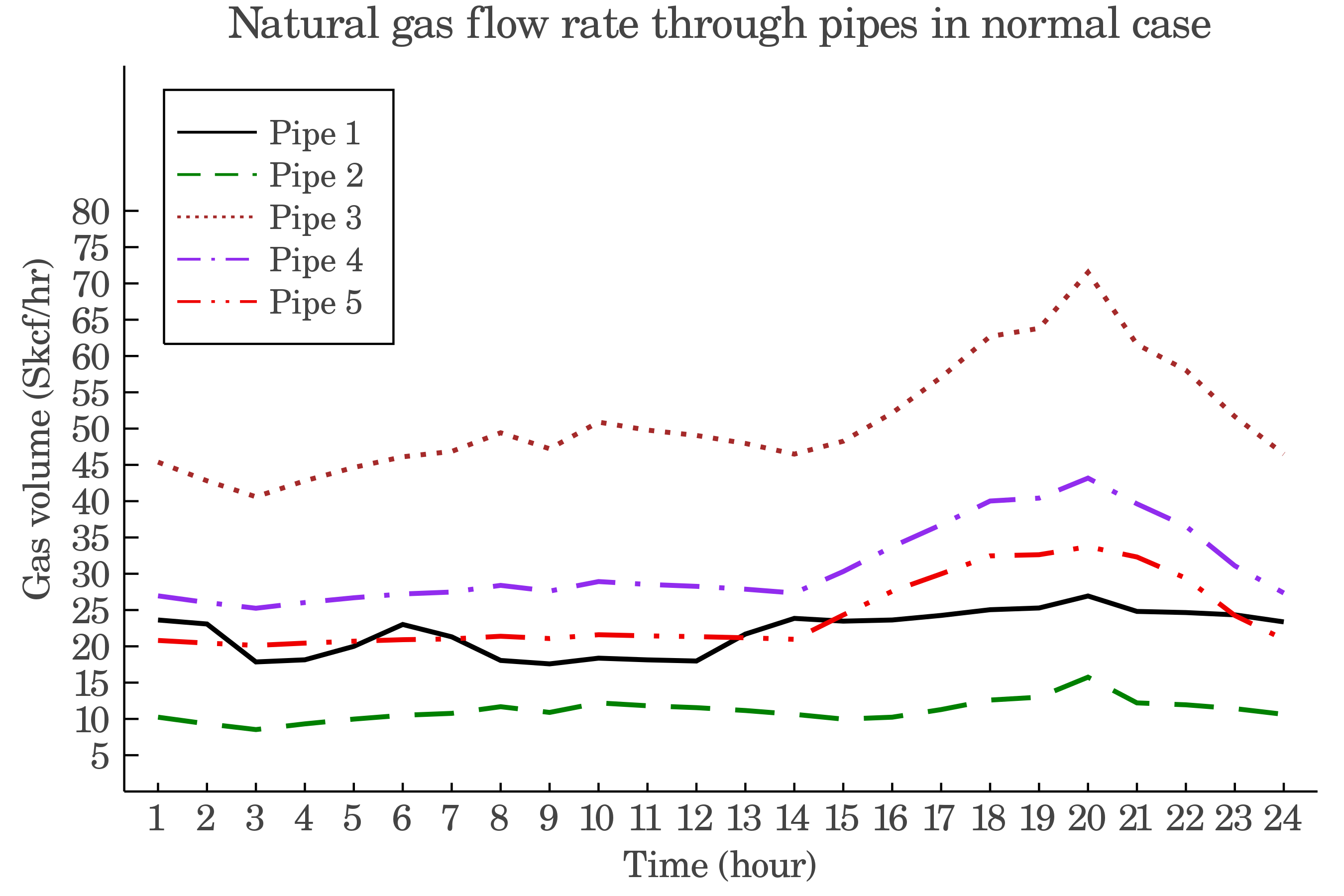}
        \vspace{-0.35cm}
    \caption{Natural gas flow of pipelines in Case 1}
    \label{fig:case3_flow_normal}
       \vspace{0.4cm}
    \includegraphics[width=6.8cm]{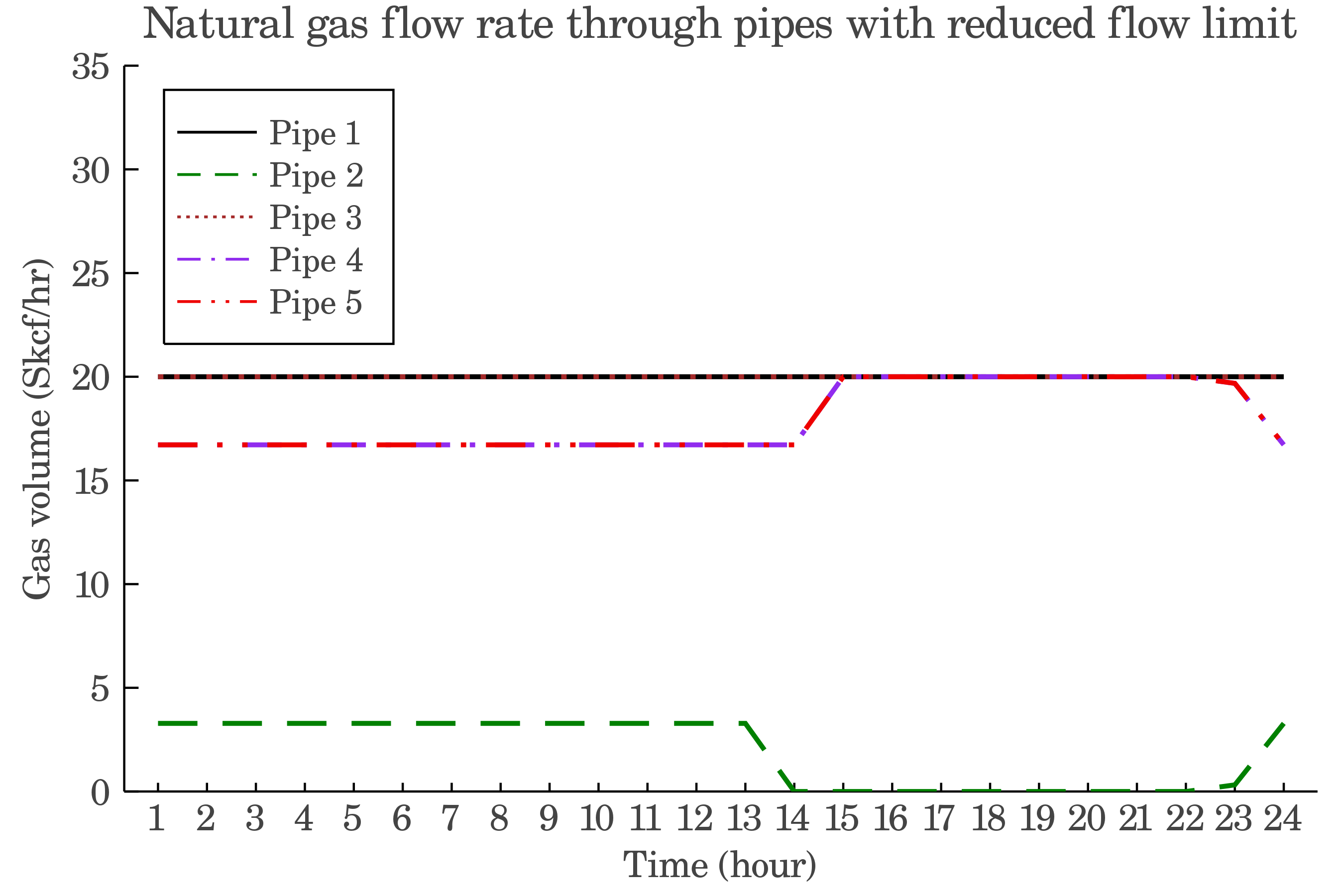}
        \vspace{-0.35cm}
    \caption{Natural gas flow of pipelines in Case 3}
    \label{fig:case3_flow_limited}       \includegraphics[width=6.8cm]{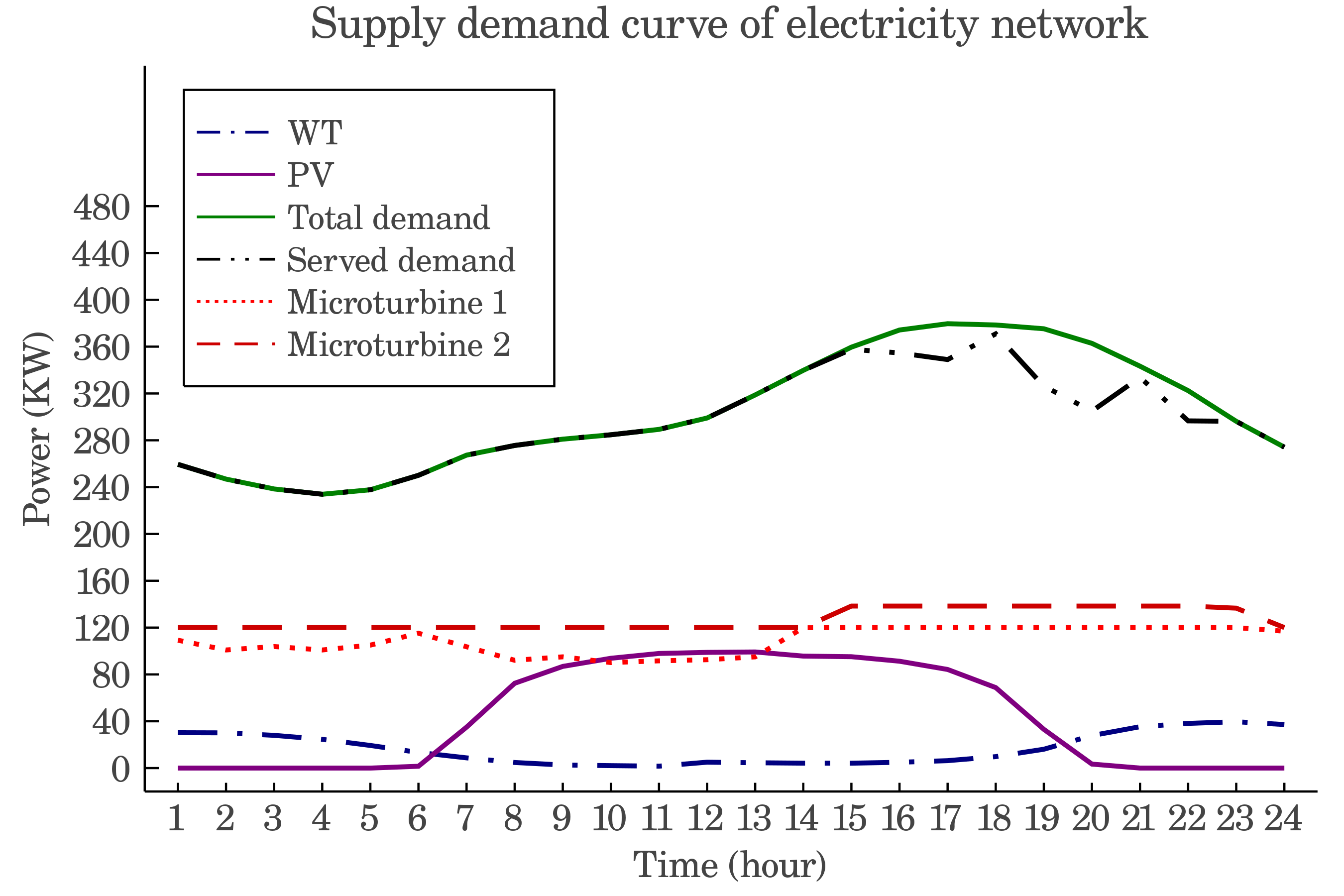}
        \vspace{-0.35cm}
    \caption{Electricity generation and consumption in Case 3}
    \label{fig:case3_elec_dispatch}
    \vspace{0.4cm}
    \includegraphics[width=6.8cm]{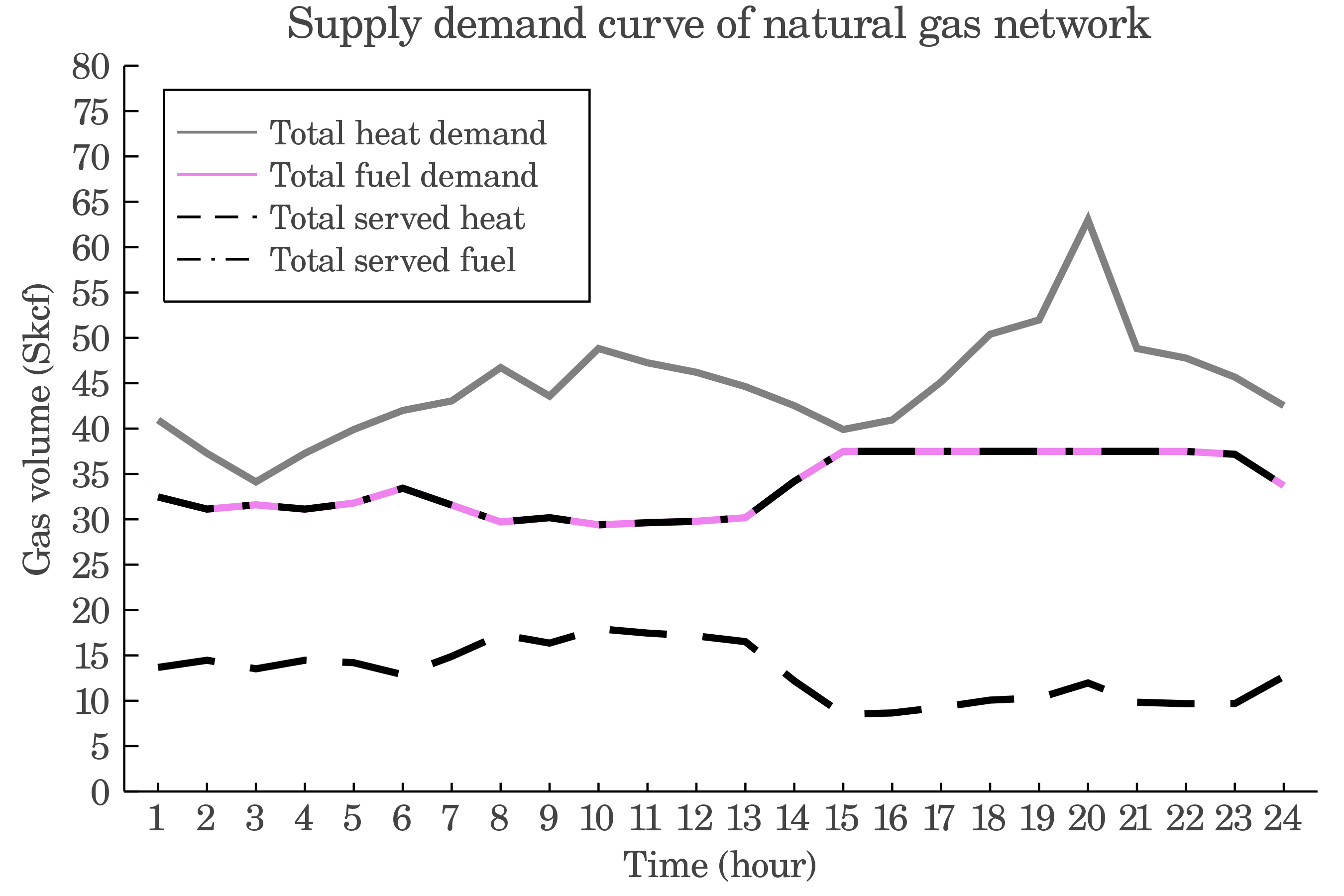}
        \vspace{-0.35cm}
    \caption{Natural gas supply demand in Case 3}
    \label{fig:case3_gas_dispatch}
    \vspace{0.4cm}
     \includegraphics[width=6.8cm]{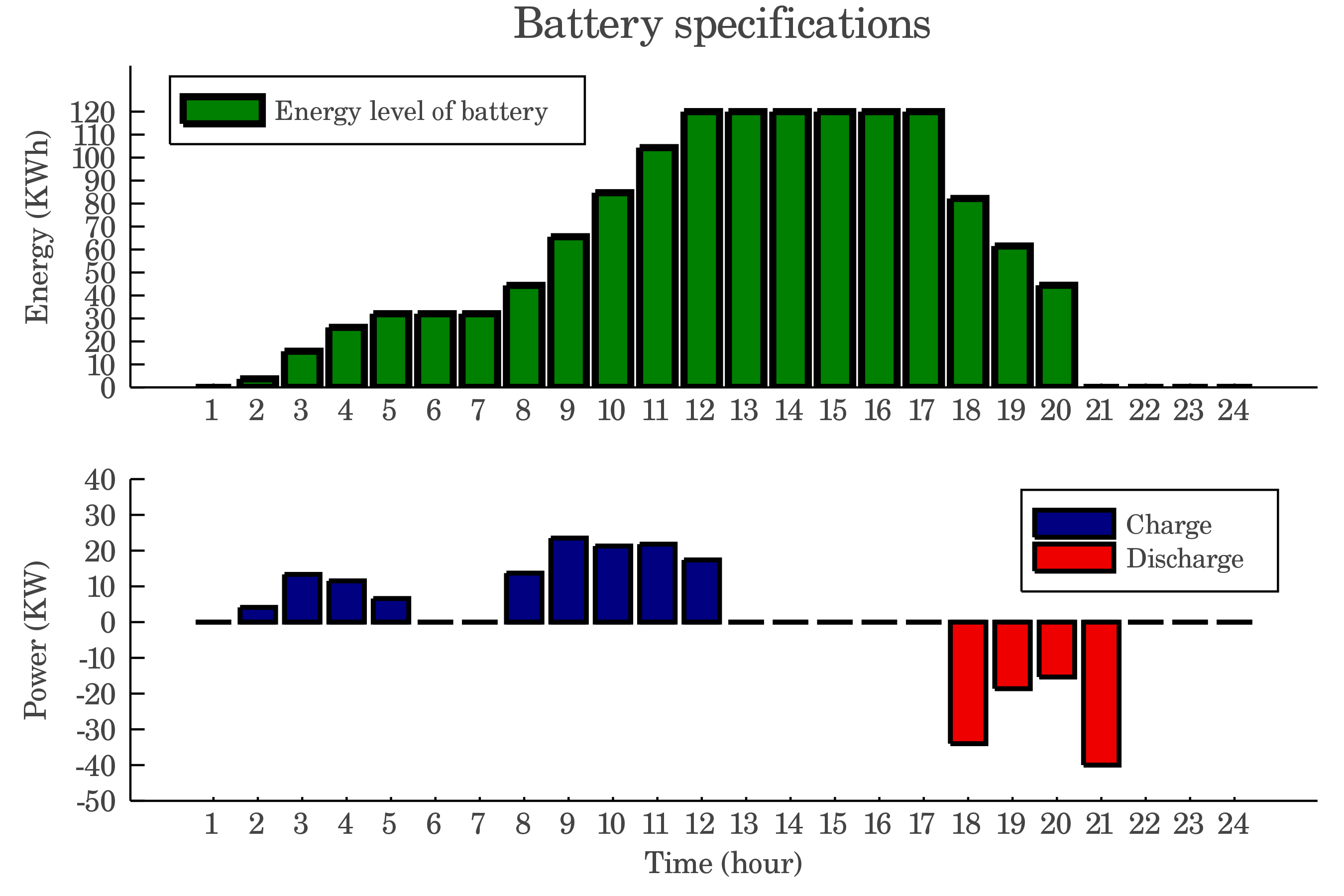}
        \vspace{-0.35cm}
    \caption{Battery energy level and power in Case 3}
    \label{fig:case3_battery}
\end{figure}

In this case, the impact of changing the natural gas pipelines maximum flow rate capacity is explored. In the normal operational conditions, this limit is set to 75 ($Skcf/hr$) (Fig. \ref{fig:case3_flow_normal}). To demonstrate the impact of this constraint on electricity network, the flow rate limit is reduced to 20 ($Skcf/hr$) in this case (Fig. \ref{fig:case3_flow_normal}). In Figures \ref{fig:case3_elec_dispatch} and \ref{fig:case3_gas_dispatch}, supply demand balance of energy in electricity and natural gas networks are displayed, respectively. It is noticed in Fig. \ref{fig:case3_elec_dispatch} that due to the limited amount of fuel provision to the microturbines, they can not operate at their maximum power rating. Consequently, 201.9 KWh of demand energy is shed throughout the day.\\
As observed from Fig. \ref{fig:case3_gas_dispatch}, only 29.2\% of the heat demand is served in this case. We consider a problem setting where the natural gas network has to supply all the fuel demand. That is why all the fuel demand is met. Nevertheless, total daily energy generation in this case is reduced from 5846 KWh (for the normal operational conditions) to 5655 KWh. The fuel cost is also reduced from \$14,432 to \$13,823. One interesting observation is the increased amount of battery utilization in this case, compared to case 1. It can be observed from Fig. \ref{fig:case3_battery} that the battery unit in this case is fully charged during the midday, when solar generation is at its peak. Later on, during hours 18-20 of the day, battery unit is fully discharged to mitigate the impact of generation deficiency. In this case, the battery degradation cost is equal to \$241.3, as opposed to the \$144.6 degradation cost in case 1. This case is a perfect example of how battery unit can help serving more demand. If there were no batteries in this system, an extra 108 KWh of demand energy were to be missed as well.
\section{Conclusion}
In this article, a short-term operational planning framework for the multiple-energy carrier hybrid AC/DC microgrid is presented. Three different operational conditions were considered for the case studies. In the first case, it was shown that the proposed structure is capable of providing all of the natural gas and electricity demand at all times, successfully. With the help of the battery storage unit, the multiple energy carrier microgrid is able to satisfy the required demand even in the peak hours, when total generations fall short of peak demand.\\
In the next two cases, the importance of inverter and pipeline capacity were demonstrated. Since most of the generation in the discussed network is injected from the DC side to the AC side, the inverters play a crucial role. If at some time of the day, this injected energy reaches the maximum capacity of the inverter, some demand is going to be missed. The flow limit in natural gas pipeline is also pivotal in providing both heat and electricity demand. In the last case, it was shown that if the pipelines' flow limit is reduced, a large portion of heat demand is going to be missed. Since the priority is with microturbine fuel, the system first feeds this type of demand. Although in this case all of the fuel demand is served, some of the electricity load is not met. This is due to the fact that the system operator is aware of the pipeline limits.
\bibliographystyle{IEEEtran}

\bibliography{conference_format.bbl}

\end{document}